\RequirePackage[2020-02-02]{latexrelease}
\documentclass[aps,prb,twocolumn,amsmath,superscriptaddress,amssymb]{revtex4}
\usepackage{graphicx}
\usepackage{enumitem}
\usepackage[colorlinks,bookmarks=false,citecolor=red,linkcolor=blue,urlcolor=blue]{hyperref}

\newcommand{\bea}{\begin{eqnarray}}          
\newcommand{\eea}{\end{eqnarray}}

\allowdisplaybreaks

\begin{document}

\title{Quantum dimer model on fullerenes: resonance, scarring and confinement}
\author{R. Krishna Prahlaadh}
\email{krishnap17@iiserb.ac.in}
\affiliation{Indian Institute of Science Education and Research (IISER), Bhopal, India}
\author{R. Ganesh}
\email{r.ganesh@brocku.ca}
\affiliation{Department of Physics, Brock University, St. Catharines, Ontario L2S 3A1, Canada}
\date{\today}

\begin{abstract}
The earliest known examples of quantum superposition were found in organic chemistry -- in molecules that `resonate' among multiple arrangements of $\pi$-bonds. Small molecules such as benzene resonate among a few bond arrangements. In contrast, a large system that stretches over a macroscopic lattice may resonate among an extensive number of configurations. This is analogous to a quantum spin liquid as described by Anderson's resonating valence bond (RVB) picture. In this article, we study the intermediate case of somewhat large molecules, the C$_{20}$ and C$_{60}$ fullerenes. We build a minimal description in terms of quantum dimer models, allowing for local resonance processes and repulsion between proximate $\pi$-bonds. This allows us to characterize ground-states, e.g., with C$_{60}$ forming a superposition of 5828 dimer covers. Despite the large number of contributing dimer covers, the ground state shows strong dimer-dimer correlations. Going beyond the ground state, the full spectrum of C$_{60}$ shows many interesting features. Its Hilbert space is bipartite, leading to spectral reflection symmetry. It has a large number of protected zero-energy states. In addition, the spectrum contains many scar-like states, corresponding to localized dimer-rearrangement-dynamics. Resonance dynamics in QDMs can manifest in the behaviour of defects, potentially binding defects via an effective attractive interaction. To test this notion, we introduce pairs of vacancies at all possible separations. Resonance energy reaches its lowest value when the vacancies are closest to one another. This suggests confinement of monomers, albeit within a finite cluster. We discuss qualitative pictures for understanding bonding in fullerenes and draw connections with results from quantum chemistry.
\end{abstract}
                                 
\maketitle
\section{Introduction}

A quantum spin liquid is a magnetic phase with no long range order and a high degree of entanglement. Within the resonating valence bond picture of Anderson\cite{Anderson1973}, it is described as a superposition of multiple singlet arrangements. Spins within the magnet pair up to form singlets, leading to a macroscopic state with no magnetization. However, pairs can be formed in many ways, corresponding to various arrangements of singlets on the underlying lattice. The entire magnet is in a coherent superposition of all such configurations. This point of view is inspired by the concept of resonance in organic chemistry, exemplified by benzene and its two Kekul\'e structures. This ultimately derives from the notion of quantum superposition which allows a system to simultaneously exist in multiple configurations. The quantum dimer model (QDM) is a framework that is specifically designed to capture these ideas\cite{RokhsarKivelson1988,Moessner2008}.

In this article, we study QDMs on fullerene graphs. As molecules containing few tens of Carbon atoms, fullerenes can exhibit resonance over mesoscopic length scales\cite{Klein1986,Flock1998}. Fullerenes also differ from typical benzenoid systems in geometry, as they form spherical surfaces. The closed, boundary-less space of a sphere may allow for a higher degree of resonance. This idea has motivated quantum chemical studies on `spherical aromaticity'\cite{Buehl2001,Chen2005}. Graph theoretical and combinatorial concepts have been employed to shed light on this question, e.g., in enumerating the number of Kekul\'e structures on C$_{60}$\cite{Klein1986,Vukicevic2005,Vukicevic2011}. We build on these results to construct a minimal statistical-physics-inspired model. As our approach neglects several microscopic details, we do not claim to describe the precise nature of chemical bonding. Nevertheless, our results provide insight into the physics of fullerenes and fullerene-based materials. They can help interpret the results of more realistic calculations as well as experimental data.

Canonical examples of quantum spin liquids are found in QDMs on macroscopic lattices\cite{MoessnerSondhi2001}. Their ground states involve resonance among an extensive number of configurations. Stationary states are divided into topological sectors, usually distinguished by winding numbers\cite{Kivelson1987}. Our results on fullerenes show analogues of these features on a finite system. In lattice systems, the behaviour of defects is known to reflect the nature of the ground state. A high degree of resonance leads to deconfined defects, with no energy cost to separate a pair of defects. Ordered ground states, with no resonance, are confined phases where defects cannot be separated. In the fullerenes, the ground state is highly resonating and yet there is an energy cost to separate defects. Admittedly, this is for finite separations within a  finite system. Nevertheless, it suggests that resonance-mediated interactions can play a role at mesoscopic length scales.

The smallest known fullerene is C$_{20}$, with twenty Carbon atoms located at the sites of a dodecahedron\cite{Lin2010}. The first fullerene to be discovered was C$_{60}$, also called Buckminsterfullerene, which forms a truncated icosahedron\cite{Kroto1985}. There are several other members of the fullerene family, with sizes reaching few thousands of Carbon atoms\cite{Hagelberg2017}. The main focus of this article is C$_{60}$. However, we also discuss C$_{20}$ as a simpler case that illustrates our approach. 

\section{Framework: dimer covers and QDM construction}

\begin{figure}
\includegraphics[width=2.2in]{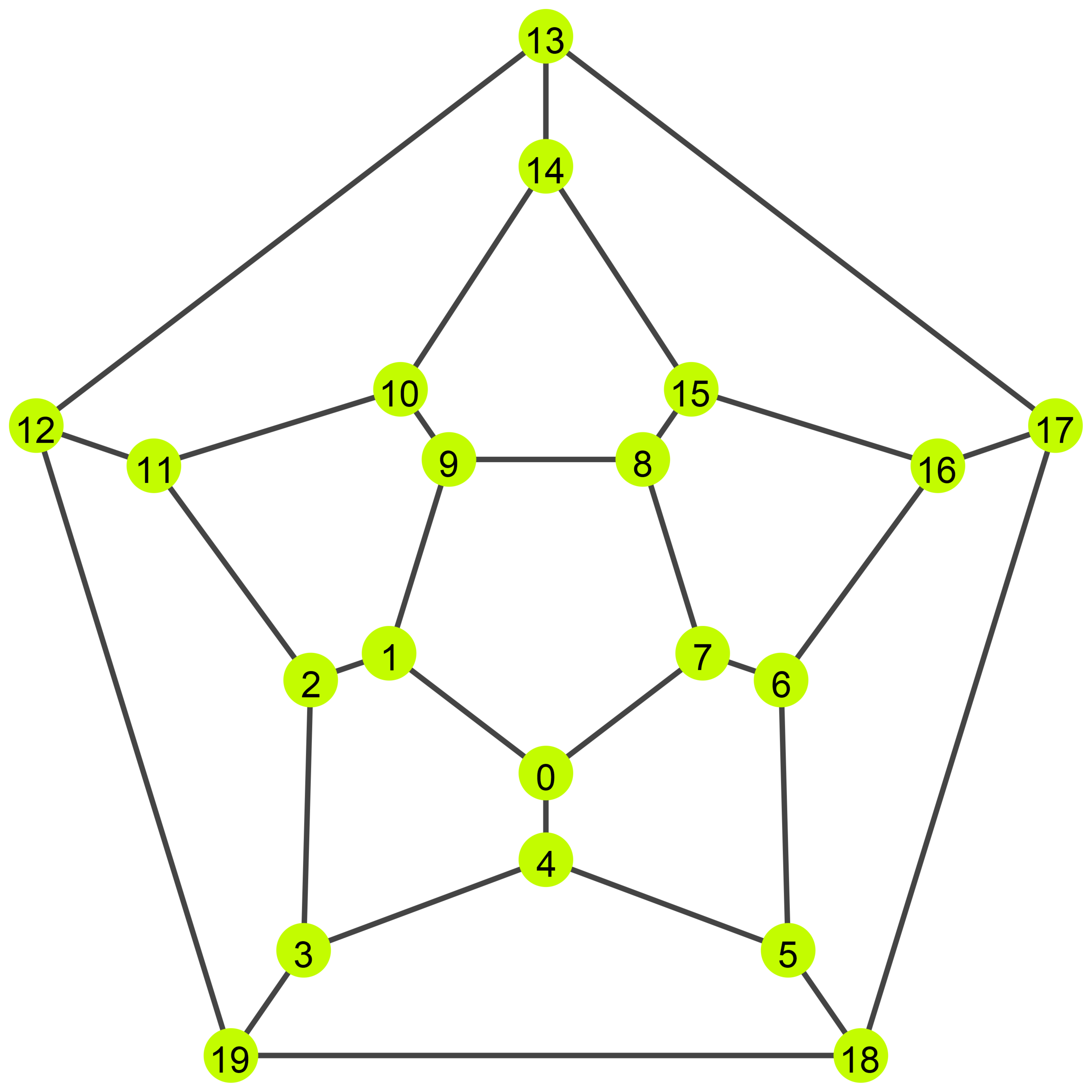}
\includegraphics[width=3in]{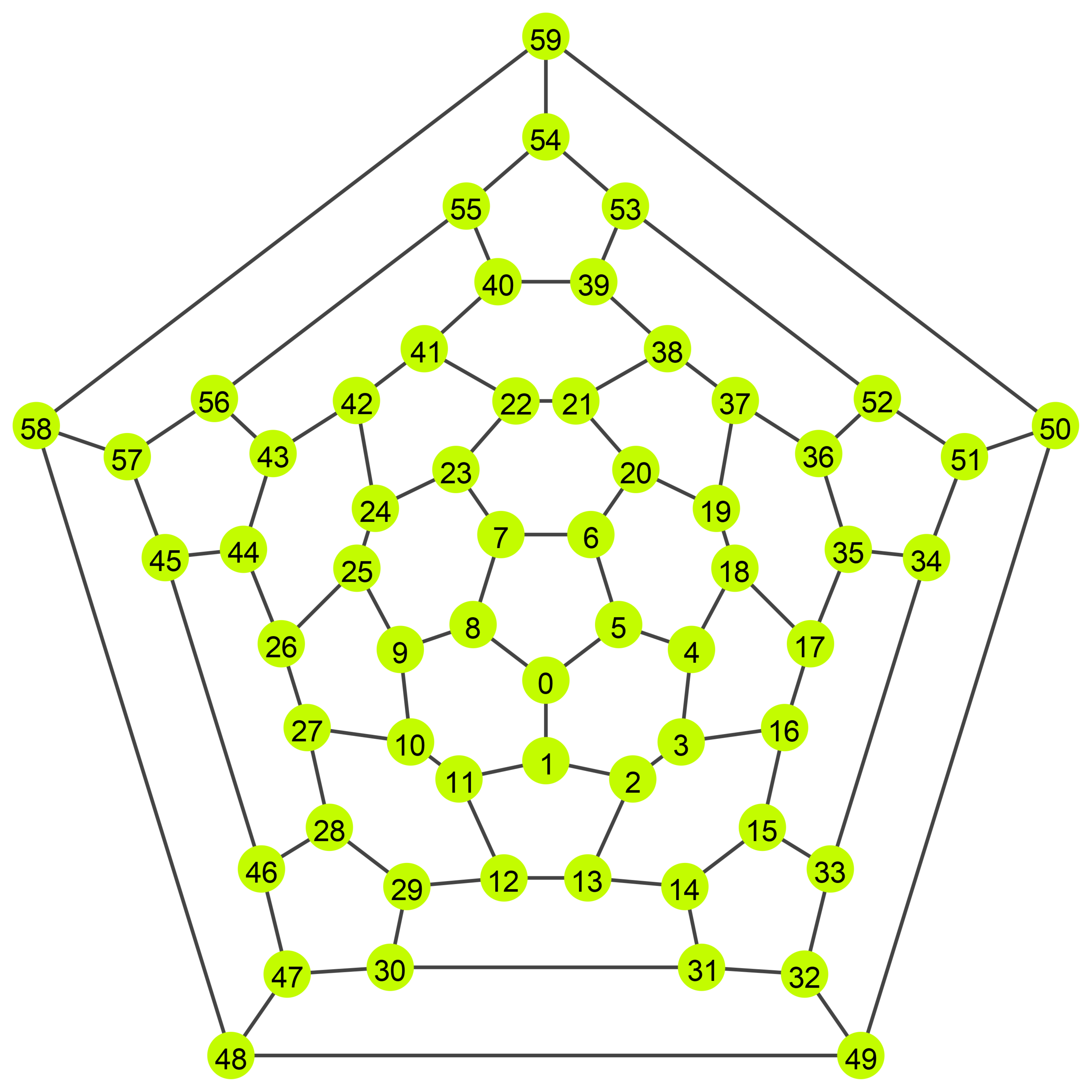}
\caption{Schlegel diagrams for C$_{20}$ (top) and C$_{60}$ (bottom). }
\label{fig.c20_schlegel}
\end{figure}

Each fullerene molecule can be viewed as a discretization of a sphere. This can be represented as a graph where the nodes are Carbon atoms, with proximate atoms connected by bonds. If these graphs are flattened onto a plane, they are called Schlegel diagrams. Fig.~\ref{fig.c20_schlegel} shows the Schlegel diagrams corresponding to C$_{20}$ and C$_{60}$.
As seen in the figure, each Carbon atom has three proximate neighbours. This is a property that is shared by all fullerenes. Ignoring the effects of curvature, we assume that each Carbon atom forms three $\sigma$-bonds with these neighbours. In order to satisfy Carbon's valency of four, each atom must form one $\pi$-bond with one of its neighbours. A configuration where every site is attached to precisely one double-bond is called a Kekul\'e structure or a dimer cover. On a graph with $N_s$ sites and $N_b$ bonds, a dimer cover will have $N_s/2$ dimers that can occupy $N_b$ locations. The dimers obey a hardcore constraint, whereby only one dimer can reside on bonds connected to a single site. In C$_{20}$, we have 10 dimers on the 30 bonds of the Schlegel diagram. In C$_{60}$, we have 30 dimers that reside on 90 bonds. 

The set of all dimer covers serves as the configuration space for the QDM. Given a graph, enumerating all dimer covers is a well-known problem. The Fisher-Kasteleyn-Temperley (FKT) algorithm\cite{Harary1967} provides an efficient method to count the number of dimer covers on any planar graph. Although C$_{20}$ and C$_{60}$ are spherical molecules, their Schlegel diagrams are planar graphs -- as seen from Fig.~\ref{fig.c20_schlegel}, the edges do not cross one another. However, the FKT algorithm does not suffice -- it only gives the total number of dimer covers, not the precise form of each dimer cover. In every calculation described below, we take the following approach. We first use an FKT routine\cite{deford_alaska} to ascertain the number of dimer covers. We then generate dimer covers using a stochastic branching algorithm, proceeding until the number of unique dimer covers matches the FKT result.

Within the QDM approach, dimer covers are taken to be orthogonal to one another, forming a Hilbert space. Dynamics arises from local rearrangements of dimers. This requires loops containing an even number of bonds. If such a loop contains alternating dimers, they can be shifted without disturbing dimers positioned elsewhere. The dominant contribution arises from the smallest such loops. Potential energy, arising from repulsion between dimers, is also taken into account. The dominant contribution stems from plaquettes that are maximally packed with dimers. Below, for C$_{20}$ and C$_{60}$, we describe the dominant kinetic and potential energy terms and include them within a Hamiltonian for the QDM.

\section{Quantum dimer model on C20}
The structure of C$_{20}$ is shown in Fig.~\ref{fig.c20_schlegel}(top) as a Schlegel diagram. The number of dimer covers on this graph is 36. We enumerate these states to characterize the Hilbert space for the QDM.

We construct the QDM Hamiltonian as follows. The smallest closed loops in C$_{20}$ are pentagons. As they have an odd number of sites, they do not allow for local dynamics. That is, dimers within a pentagon cannot be rearranged without disrupting the bond configuration elsewhere. The next largest loops contain eight sites, enclosing two neighbouring pentagons. If one such eight-site loop hosts four dimers on alternating bonds, this allows for a local re-arrangement where each dimer can be shifted by one bond along the loop. We assign a `hopping amplitude', $t$, to this process. We next consider repulsion between dimers. On a given pentagon, we may have zero, one or two dimers. We associate a repulsive energy cost, $V$, with every pentagon that hosts precisely two dimers. These considerations lead to the following Hamiltonian,
\bea
H = -t\sum_{\includegraphics[width=0.2in]{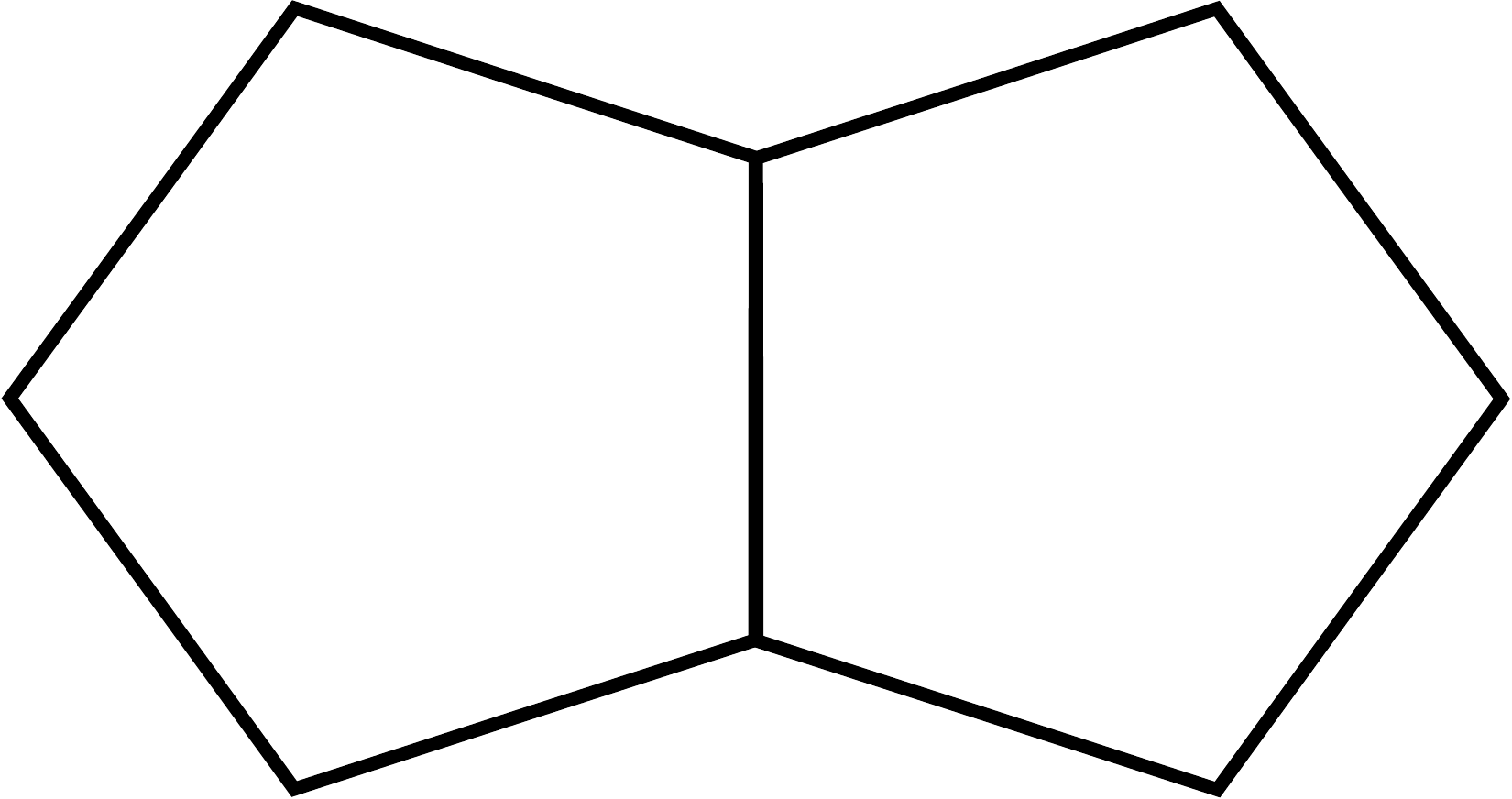}}\Big\{ \big\vert  \begin{gathered}\includegraphics[width=0.3in]{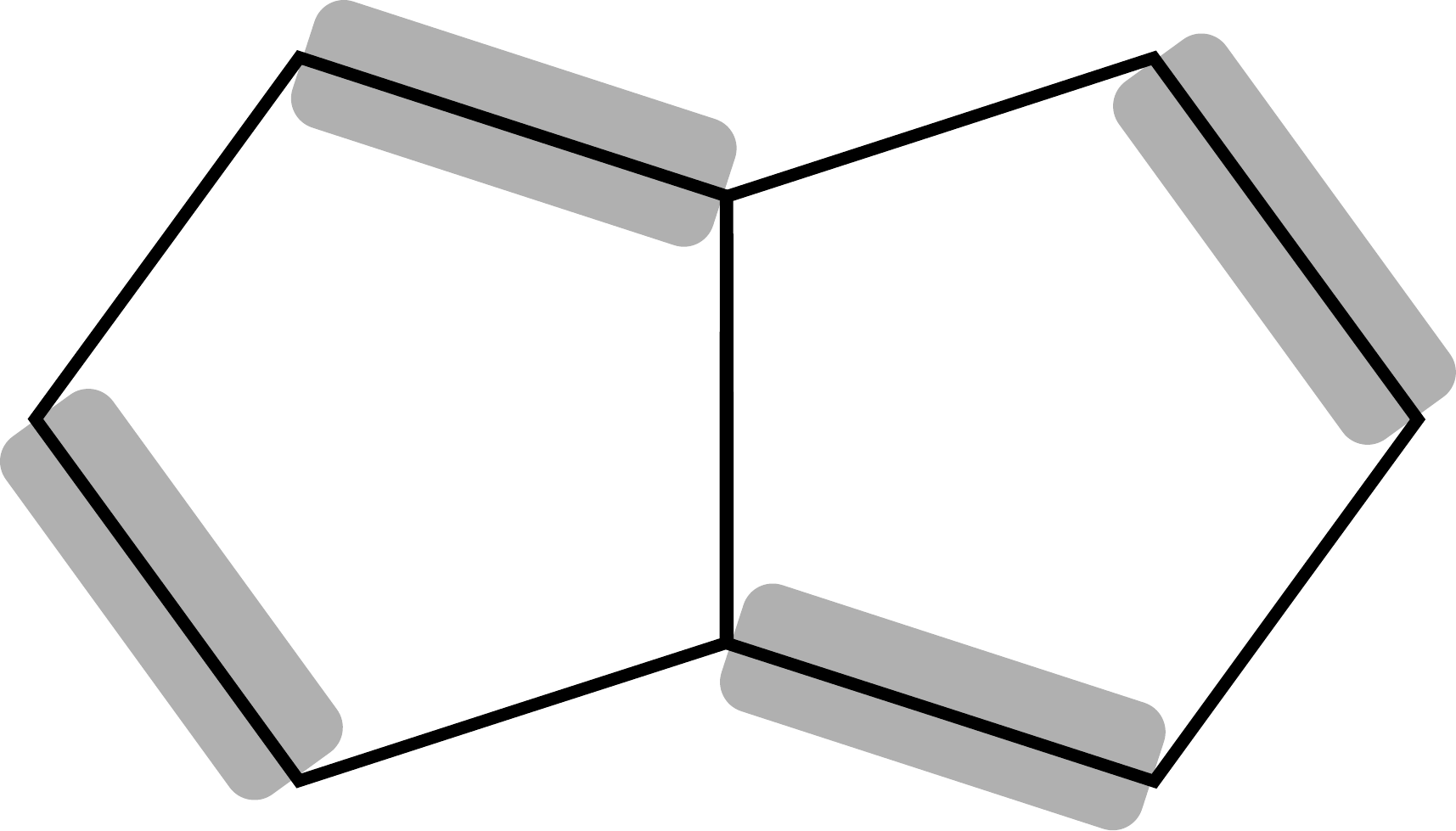} \end{gathered}\big\rangle \big\langle \begin{gathered} \includegraphics[width=0.3in]{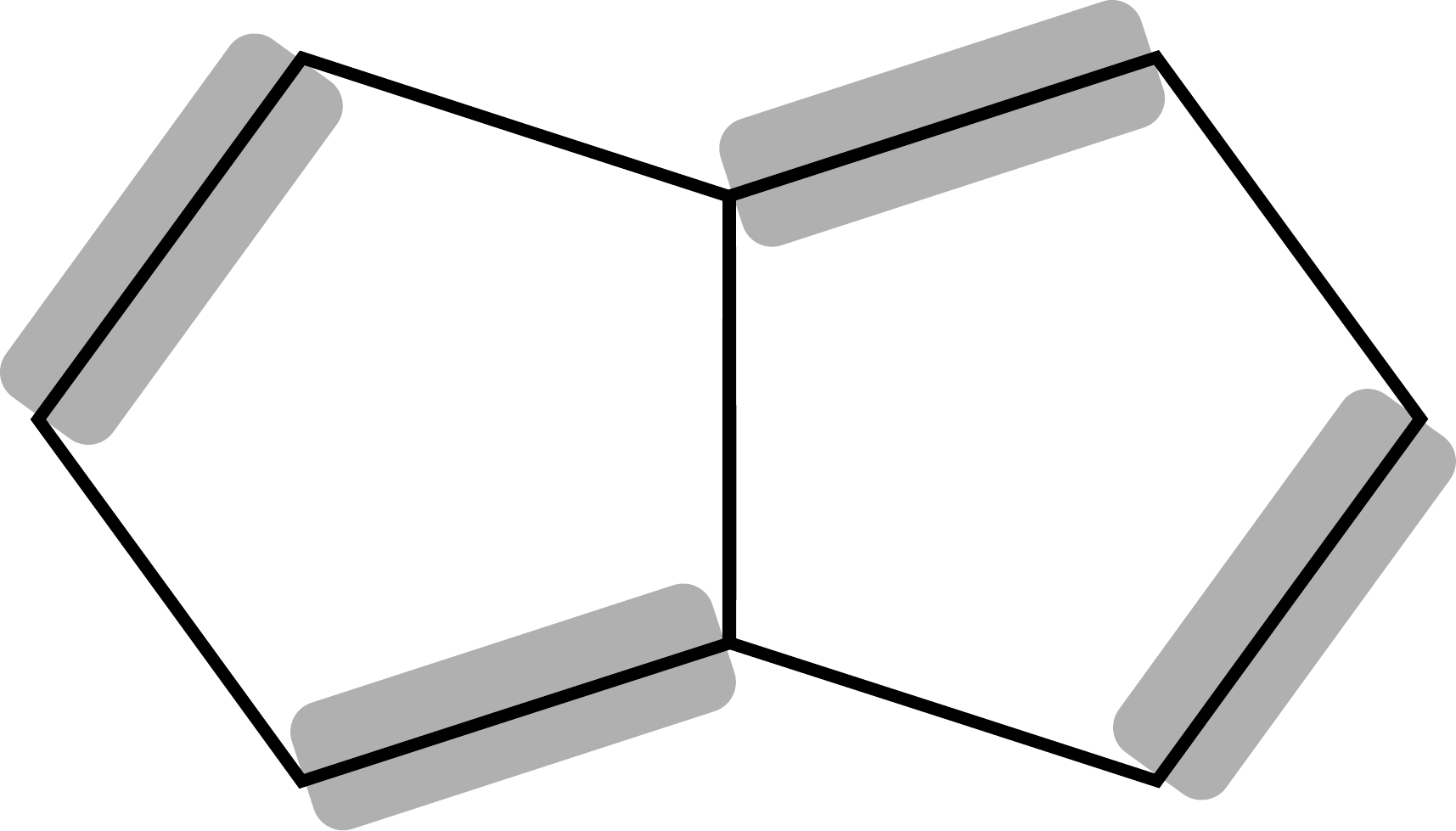} \end{gathered}\big\vert + h.c. \Big\}+ V \sum_{ \includegraphics[width=0.1in]{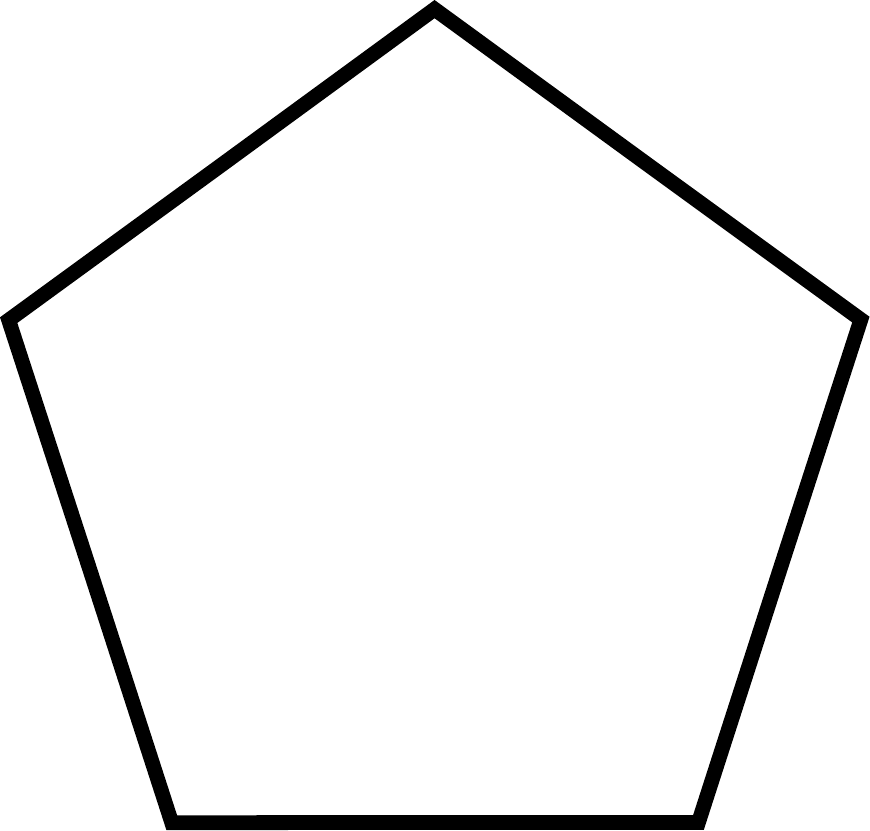} }\delta(\hat{n}_{\includegraphics[width=0.1in]{pent} } - 2),~
\eea
where the sum over $\includegraphics[width=0.2in]{2pent}$ runs over all pairs of neighbouring pentagons. In the potential term, the sum runs over every pentagon of C$_{20}$. The operator $\hat{n}_{\includegraphics[width=0.1in]{pent} } $ counts the number of dimers residing on a given pentagon. If the number is two, the system incurs a repulsion cost of $V$.

\subsection{Ground states in the undoped model}
In the basis of dimer covers, the Hamiltonian takes the form of a 36 $\times$ 36 matrix. We obtain the eigenspectrum by diagonalizing this matrix. A particular feature of this Hamiltonian is its full connectivity. Any element of the Hilbert space can be reached from any other element, by repeated application of the Hamiltonian.  

\begin{figure}
\includegraphics[width=2in]{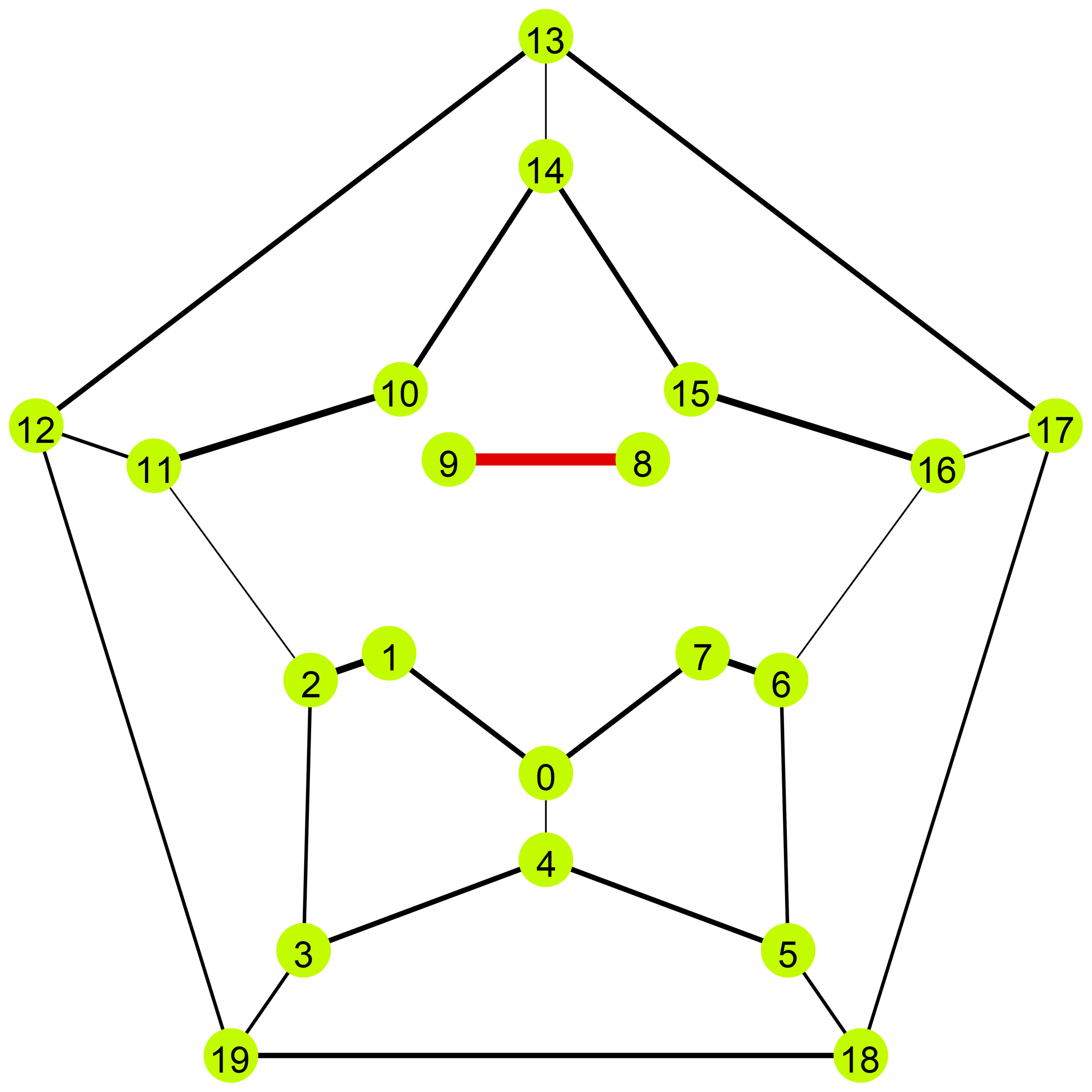}
\caption{Bond correlations in the ground state of C$_{20}$ when $V/t=0$. The reference bond is shown in red. On all other bonds, the bond thickness shown is proportional to the correlation with respect to the reference bond (see text). }
\label{fig.C20_gs_corr}
\end{figure}
We focus on the character of the lowest energy state. For $t,V >0 $, we find a non-degenerate ground state. It retains the same qualitative character as $V/t$ is varied. To understand the ground state, we may evaluate the probability of finding a dimer on any given bond. However, as all bonds in C$_{20}$ are equivalent, the probability is uniform. We then evaluate dimer-dimer correlations with the result shown in Fig.~\ref{fig.C20_gs_corr}. In order to define this quantity, we employ a projection operator, $\hat{P}_{(i,j)}$, where $(i,j)$ represents a nearest-neighbour bond on the C$_{20}$ graph. When acting on a dimer cover, $\hat{P}_{(i,j)}$ is unity if a dimer is present at $(i,j)$ and zero otherwise. We define the dimer-dimer correlation in a state $\vert \psi \rangle$ as
\bea
p_{(i,j),(k,l)} = \langle \psi \vert \hat{P}_{(i,j)} \hat{P}_{(k,l)} \vert \psi \rangle,
\eea
yielding a real quantity that ranges from 0 to 1. This represents the probability that both bonds are simultaneously occupied by dimers. In Fig.~\ref{fig.C20_gs_corr}, the state $\vert \psi \rangle$ is chosen to be the ground state of the QDM Hamiltonian with $V=0$. The correlation is calculated with respect to a fixed reference bond, shown in red. At short distances, $p_{(i,j),(k.l)}$ shows a strong pattern -- arising from the condition that two dimers cannot touch at a site. Some correlations survive even at the furthest distance. As $V$ is increased from zero, the dimer-dimer correlation does not change significantly. The ground state itself evolves smoothly, with the first excited state always separated by a gap.

We note that kinetic energy and potential energy, as defined on C$_{20}$ here, are not truly independent. For a given dimer cover, the strength of kinetic energy can be gauged from the number of `flippable' loops. The dimer covers of C$_{20}$ fall under two classes: the first consisting of dimer covers with six flippable loops and the second with ten flippable loops. Within each class, the repulsive potential energy takes the same value. This reveals that kinetic energy and potential energy reflect the same underlying information.

\subsection{Ground states with doping}
\begin{figure}
\includegraphics[width=3.4in]{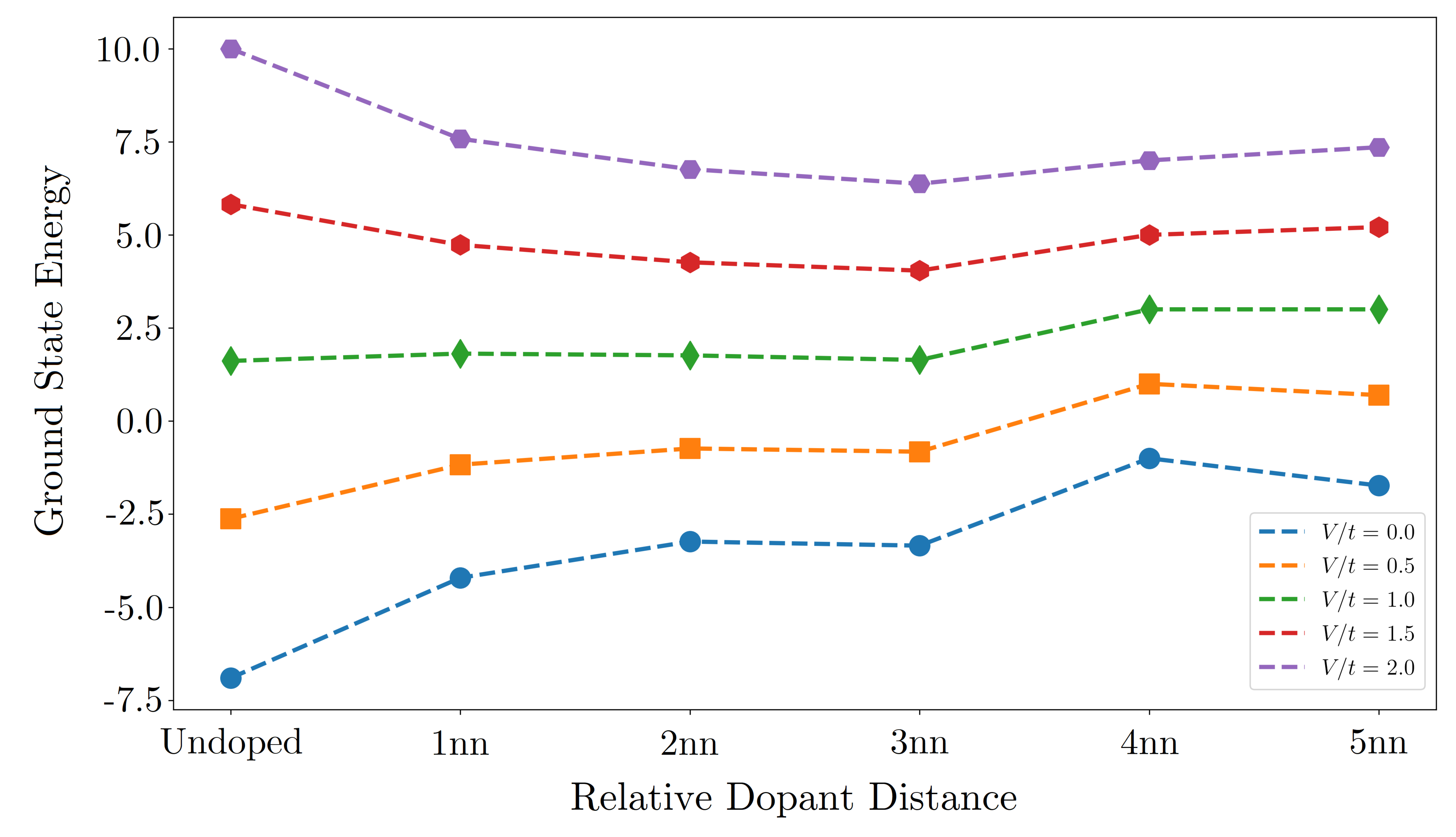}
\caption{Ground state energy vs. distance between vacancies. `Undoped' represents the C$_{20}$ graph with no vacancies. The remaining columns represent vacancies at various distances, e.g., `1nn' represents vacancies on sites that are first neighbours of one another -- say on sites 0 and 1 as shown in Fig.~\ref{fig.c20_schlegel}. }
\label{fig.C20vacancies}
\end{figure}

We next introduce a pair of static vacancies on the C$_{20}$ graph. Within the quantum dimer model, we may only introduce an even number of defects. An odd number does not allow for the remaining sites to host dimers in a consistent fashion. With a pair of vacancies, we find ground state properties for all possible relative distances between them. In Fig.~\ref{fig.C20vacancies}, we plot the ground state energy as a function of relative distance between vacancies. 

When $V/t$ is small, introducing a pair of vacancies imposes a cost. For any separation between the vacancies, the ground state energy is higher as compared to `undoped' C$_{20}$. As a function of separation, we see an overall upward trend.  
Energy is lowest when the vacancies are immediately adjacent to each other. We interpret these findings as follows. Dimers must necessarily avoid the neighbourhood of a vacancy. With two dimers, we have two inaccessible regions that cannot be accessed by dimers. This reduces the degree of resonance, manifesting as an energy cost for introducing vacancies. 
If the vacancies are adjacent to one another, the two inaccessible regions overlap to give rise to a smaller forbidden area. This allows for increased resonance with dimers moving over a larger region in the graph. This can be viewed as an effective attraction between vacancies, generated by resonance in the background.

For large $V/t$, introducing vacancies lowers the energy of the system. A pair of vacancies reduces the number of dimers by one. This allows the remaining dimers to move away from one another. This reduces the potential energy and thereby the overall ground state energy. Upon varying the separation between vacancies, the ground state energy does not vary significantly. For any separation, the system finds a similar configuration that minimizes repulsion.

\section{Quantum dimer model on C60}
Compared to C$_{20}$, the C$_{60}$ structure allows for a simpler form of dynamics. The smallest flippable loops are single hexagons that carry three dimers. A hexagon can be `flipped' if it contains three alternating dimers -- changing it from one Kekul\'e configuration to another. This represents the dominant contribution to the kinetic energy term, given by 
\bea
\hat{H}_{K.E.} = -t\sum_{\includegraphics[width=0.1in]{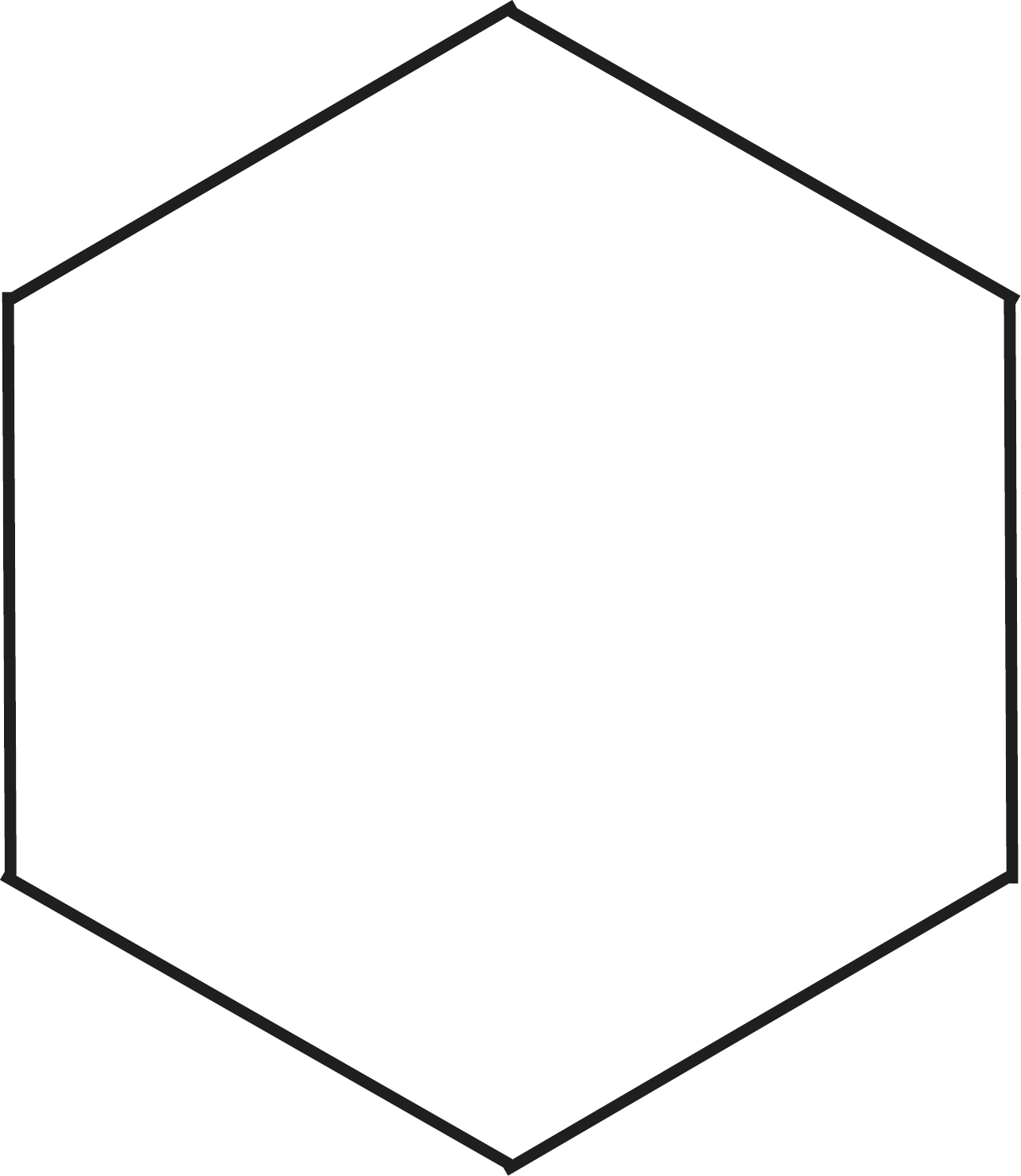}}\Big\{ \big\vert  \begin{gathered}\includegraphics[width=0.15in]{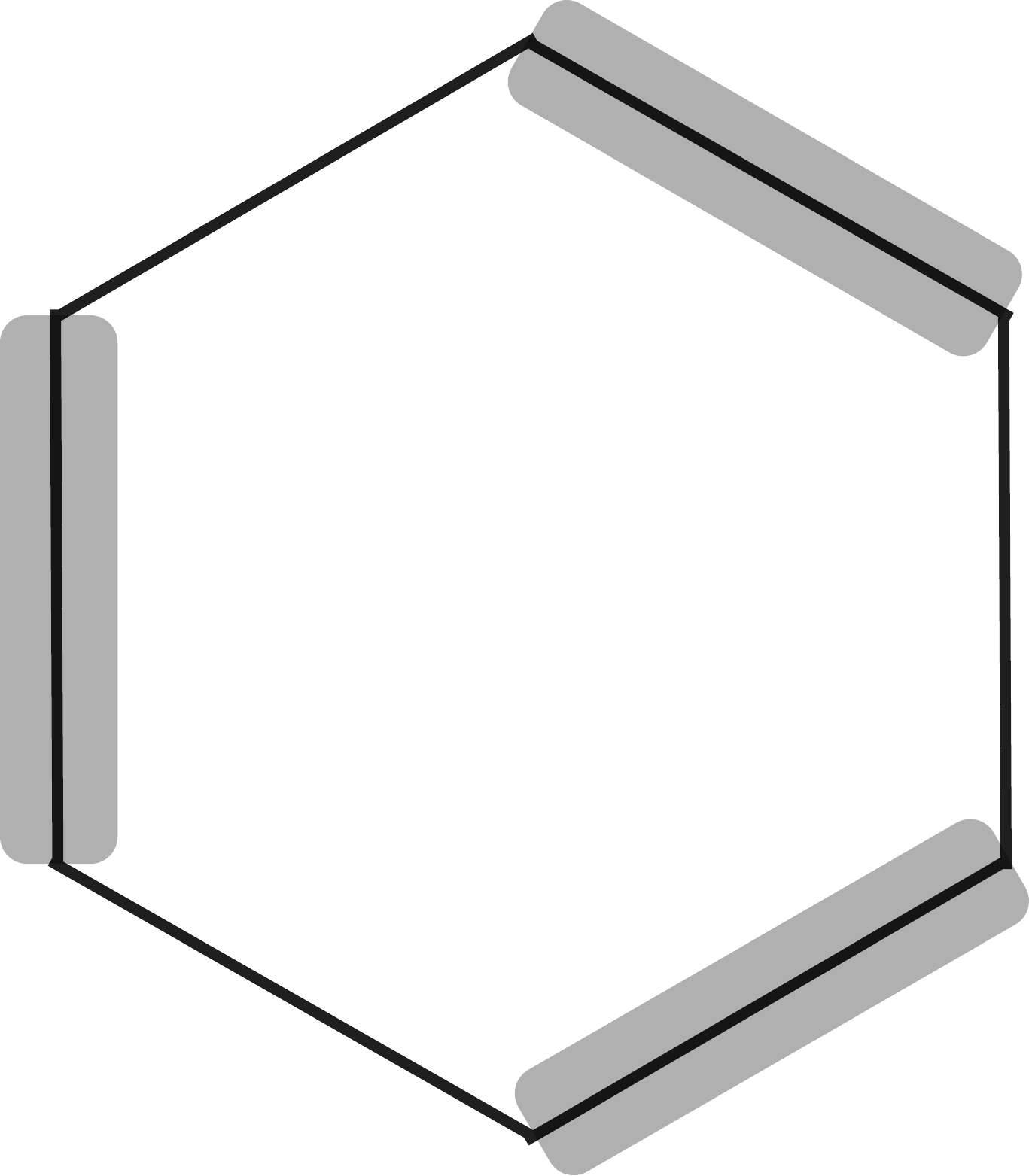}\end{gathered} \big\rangle \big\langle  \begin{gathered}\includegraphics[width=0.15in]{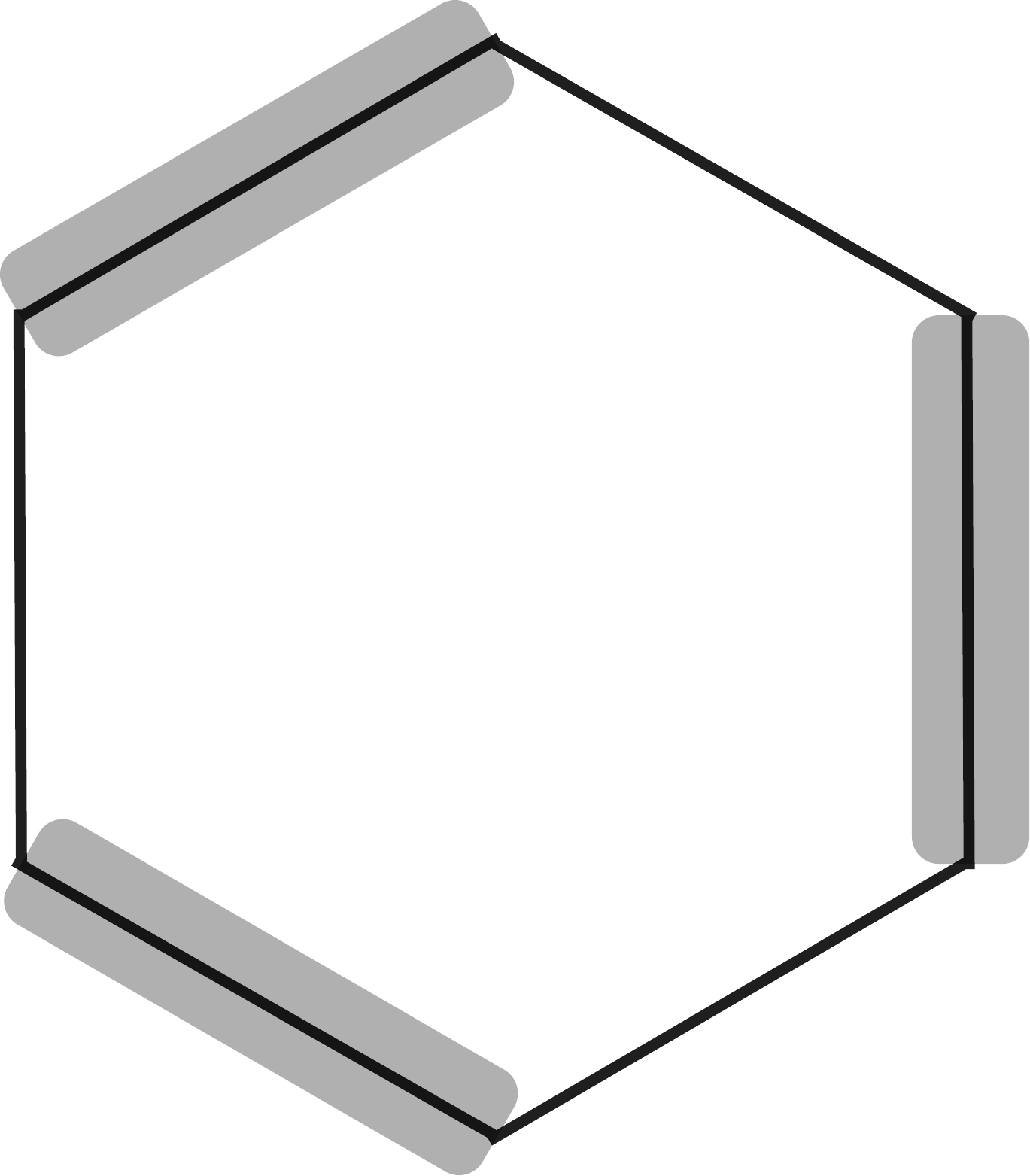} \end{gathered}\big\vert + h.c. \Big\},
\label{eq.HKE}
\eea
where the sum runs over every hexagon. We neglect weaker contributions that can arise from larger loops. 

To characterize the potential energy of repulsion, we consider the two smallest closed loops: pentagons and hexagons. A pentagon can carry at most two dimers, however, a hexagon can host up to three. We assume that the leading contribution to potential energy occurs from hexagons with three dimers. We retain this contribution and neglect all others, for simplicity. This leads to
\bea
\hat{H}_{P.E.} =V \sum_{\begin{gathered} \includegraphics[width=0.1in]{hexa} \end{gathered}} \Big[ \big\vert  \begin{gathered}\includegraphics[width=0.15in]{hexa_A} \end{gathered}
\big\rangle \big\langle \begin{gathered} \includegraphics[width=0.15in]{hexa_A}\end{gathered} \big\vert +  \big\vert  \begin{gathered}\includegraphics[width=0.15in]{hexa_B} \end{gathered}\big\rangle \big\langle \begin{gathered} \includegraphics[width=0.15in]{hexa_B} \end{gathered}\big\vert \Big],
\label{eq.HPE}
\eea
As with $\hat{H}_{K.E.}$, the sum in $\hat{H}_{P.E.}$ runs over every hexagon on the C$_{60}$ graph.

\section{The Hilbert space and its connectivity}
\begin{figure}
\includegraphics[width=3.1in]{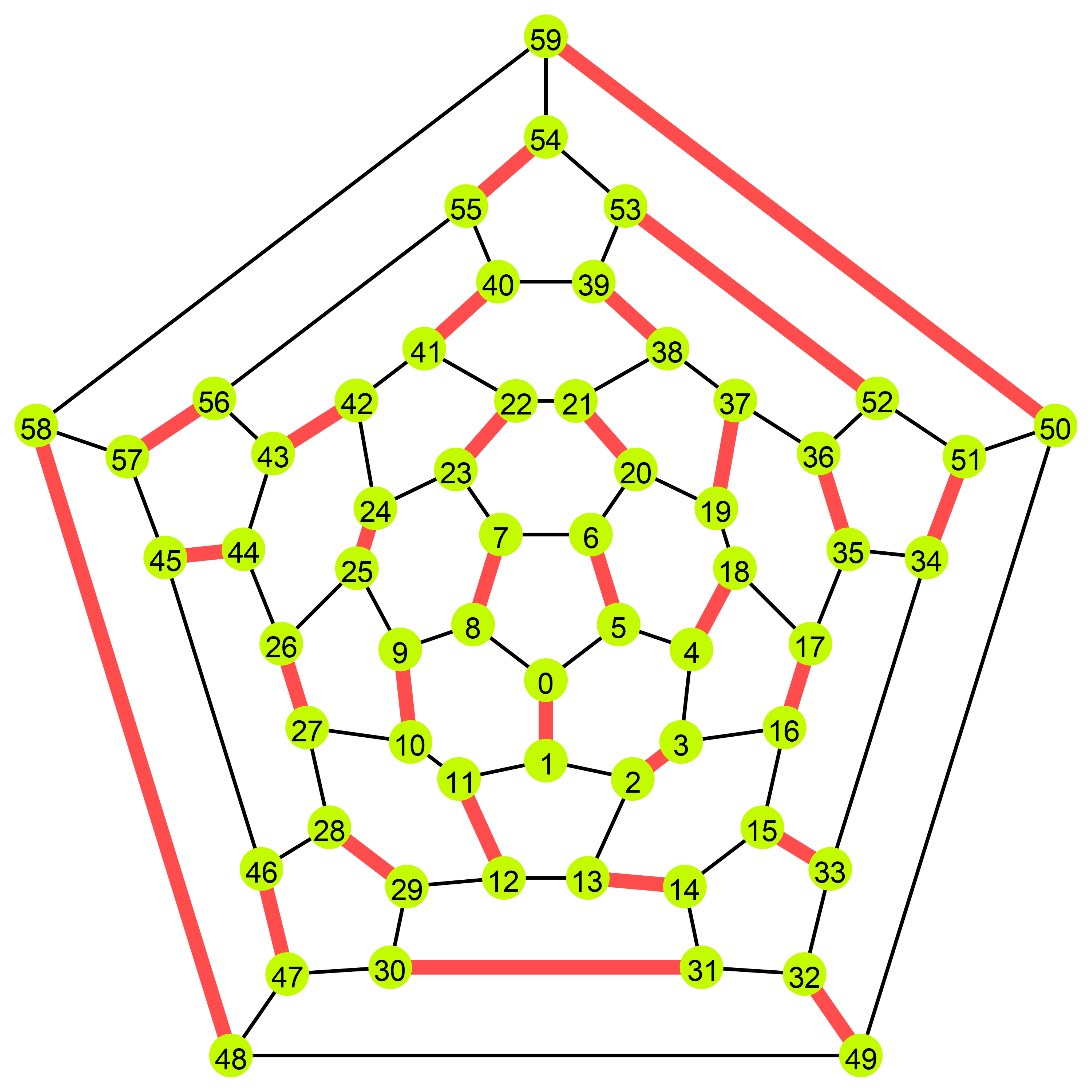}
\includegraphics[width=3.1in]{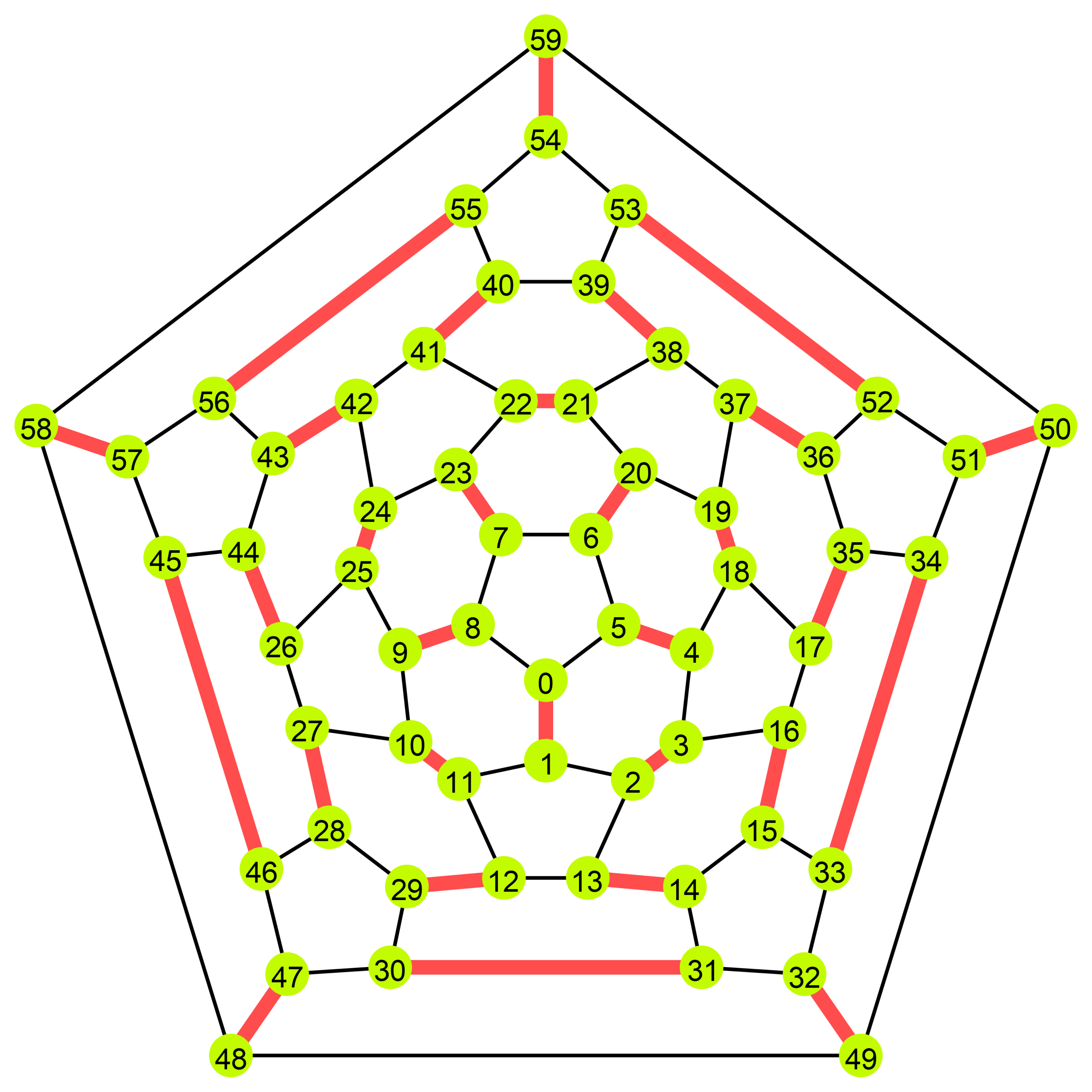}
\caption{Two dimer covers on the C$_{60}$ graph. Top: A configuration with no flippable loops. Bottom: The dimer cover with the largest number of flippable loops. Each of the twenty hexagons carries three alternating dimers.}
\label{fig.C60flippable}
\end{figure}

The Hilbert space is the set of all dimer covers on the C$_{60}$ graph. Fig.~\ref{fig.C60flippable} shows two examples. Both have thirty dimers placed on the C$_{60}$ graph, with each site attached to precisely one dimer. The total number of dimer covers is 12500\cite{Klein1986,Tesler1993}. A detailed analysis of these dimer covers and their symmetries can be found in Refs.~\onlinecite{Vukicevic2005,Vukicevic2011}.

In the basis of dimer covers, the potential energy is purely diagonal. To understand the nature of dynamics, we focus on the kinetic energy first, representing $\hat{H}_{K.E.}$  as a 12500$\times$12500 matrix. This can be viewed as dynamics on a network of 12500 nodes, with each node representing a dimer cover. Each non-zero entry of $\hat{H}_{K.E.}$  serves as a link that connects two nodes. Notably, this network is bipartite. This can be seen by characterizing each dimer cover by $N_{pent.}$, the number of dimers that reside on pentagons. A single operation of $\hat{H}_{K.E.}$, corresponding to flipping a single hexagon, changes $N_{pent.}$ by three. This follows from the fact that each hexagon in C$_{60}$ is surrounded by three pentagons. As a consequence, the Hilbert space separates into two sectors, corresponding to even and odd values of $N_{pent.}$. Under the action of $\hat{H}_{K.E.}$, a state with an even (odd) value of $N_{pent.}$ either vanishes or moves to a state with an odd (even) value of $N_{pent.}$. 

\begin{figure}
\includegraphics[width=3.3in]{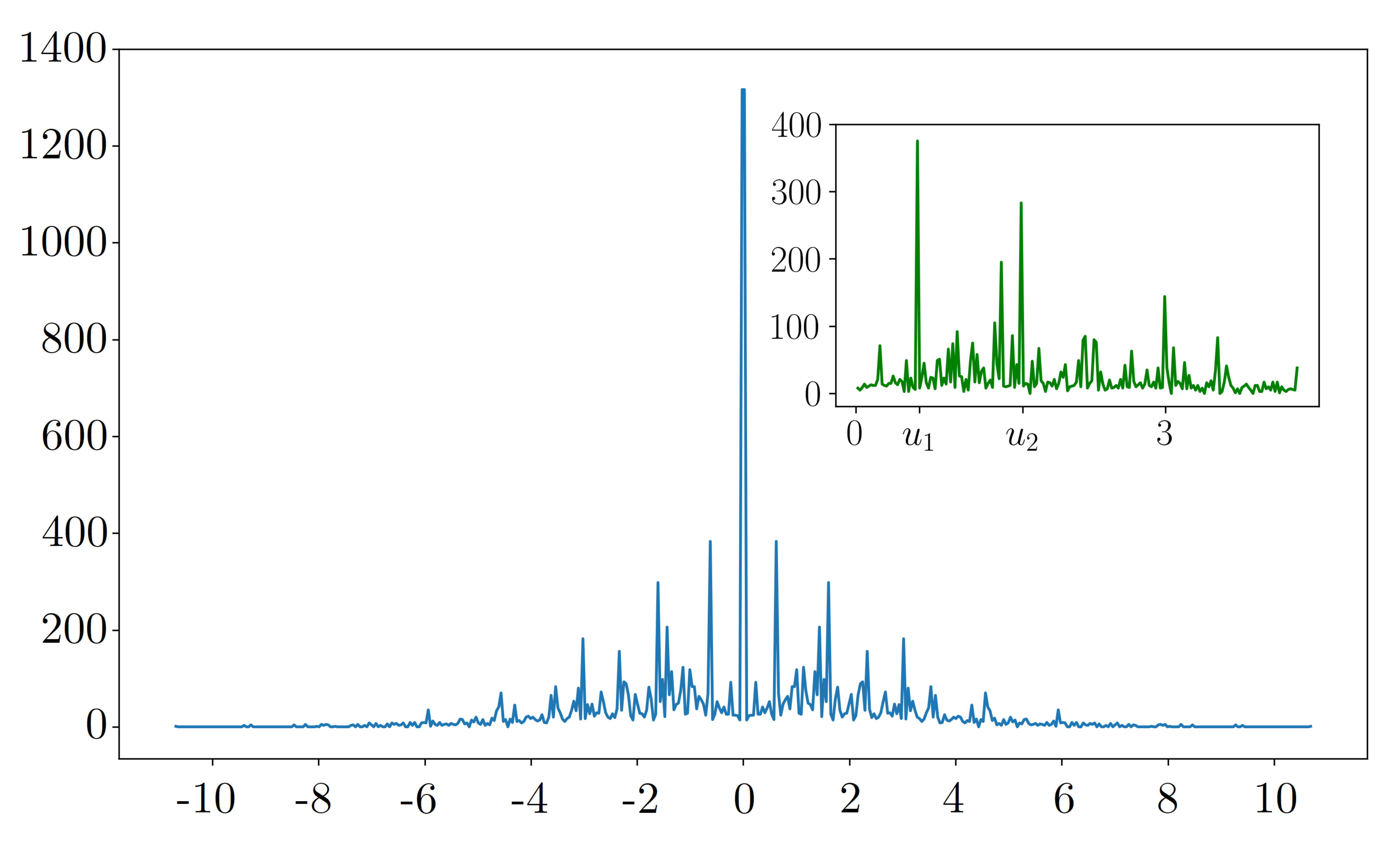}
\caption{Density of states in the C$_{60}$ QDM in the purely kinetic regime, i.e., with $V=0$. The inset shows a limited energy range where three peaks are clearly discernible. The peaks occur at $u_1 = 2 \cos(2\pi/5)$, $u_2 = 2 \cos(\pi/5)$ and $3$.}
\label{fig.dos}
\end{figure}

We formally describe this in terms of an operator, 
\bea
\hat{C} = \exp \Big( i \pi \sum_{ \includegraphics[width=0.1in]{pent}} \hat{n}_{ \includegraphics[width=0.1in]{pent} } \Big) ,
\eea
where the sum runs over the twelve pentagons in the C$_{60}$ graph. This operator returns $\pm 1$ when acting on a dimer cover, depending on whether the dimer count on pentagons is even/odd. We have
$ \hat{H}_{K.E.}\hat{C}+\hat{C}\hat{H}_{K.E.} = 0$, i.e., $\hat{C}$ anticommutes with the Hamiltonian. This immediately leads to spectral reflection symmetry\cite{Iadecola2018}. Given any eigenstate of $\hat{H}_{K.E.}$, denoted as $\vert\psi_E\rangle$ with eigenvalue $E$, we can construct another eigenstate $\vert \psi_{-E} \rangle \equiv 
\hat{C} \vert \psi_E \rangle$, with eigenvalue $(-E)$. If $E\neq 0$, these two states must necessarily be orthogonal. In other words, non-zero energy eigenvalues occur in pairs of the form $(E,-E)$. This reflection symmetry can be seen in the density of states, plotted in Fig.~\ref{fig.dos}. 

The operator $\hat{C}$ protects zero-energy states, in a manner similar to an index theorem\cite{Iadecola2018}. The quantity $W = \vert \mathrm{Tr} \{\hat{C}\} \vert$ provides a lower bound for the number of zero-energy states. This can be seen as follows. As $\hat{C}$ squares to one, its eigenvalues are $\pm 1$. The trace in $W$ can be evaluated by going to the basis of $\hat{H}_{K.E.} $-eigenstates. States with non-zero energies do not contribute to the trace as $\hat{C}$ takes each such state to an orthogonal partner. In the zero-energy sector, states can be re-expressed as eigenstates of $\hat{C}$. The trace contains contributions from states with positive and negative eigenvalues of $\hat{C}$. A non-zero value of $W$ imposes a lower bound on the number of states within the zero-energy sector. 

In C$_{60}$, we evaluate $W$ by tracing over the 12500 dimer covers to find $W=116$. The actual number of zero-energy states far exceeds this lower bound. In the $V=0$ limit, we have 1308 zero-energy states. Remarkably, as $V$ moves away from zero, we have 256 states that remain pinned at zero energy. These states are single dimer covers that are `unflippable', with no hexagon containing three alternating dimers. An example is shown in Fig.~\ref{fig.C60flippable}(top). These dimer covers are eigenstates of $\hat{H}_{P.E.}$ as well as $\hat{H}_{K.E.}$, with both having eigenvalue zero.

Under the dynamics encoded in Eq.~\ref{eq.HKE}, the Hilbert space is not fully connected. In fact, it separates into 583 independent sectors. We discuss specific sectors below.

\subsection{Maximally flippable sector}
\label{sec.maxflip}
The two dimer covers of Fig.~\ref{fig.C60flippable} represent two extremes. The configuration at the top is unflippable, as no hexagon has three alternating dimers. In fact, there are 256 such dimer covers on the C$_{60}$ graph. In contrast, at the bottom, we have a maximally flippable configuration. Each of the 20 hexagons of the C$_{60}$ graph has three alternating dimers. This configuration is unique -- all other dimer covers have fewer flippable hexagons. 

The maximally flippable dimer cover has the highest degree of connectivity. As we have 20 flippable hexagons, $\hat{H}_{K.E.}$ connects this state connects to 20 others. These states, in turn, connect to sixteen new elements each. Proceeding on these lines, we span a set of 5828 dimer covers -- all accessible from the maximally flippable configuration of Fig.~\ref{fig.C60flippable}(bottom). Taking the Hamiltonian to be a network in Hilbert space, these states form the largest connected sector. This sector contains the ground state of $\hat{H}_{K.E.}$ as it allows for the largest spread of the wavefunction. We discuss the ground state and its characteristics in Sec.~\ref{sec.gs} below.

\subsection{Scar-like sectors}
\begin{figure*}
\includegraphics[width=6.5in]{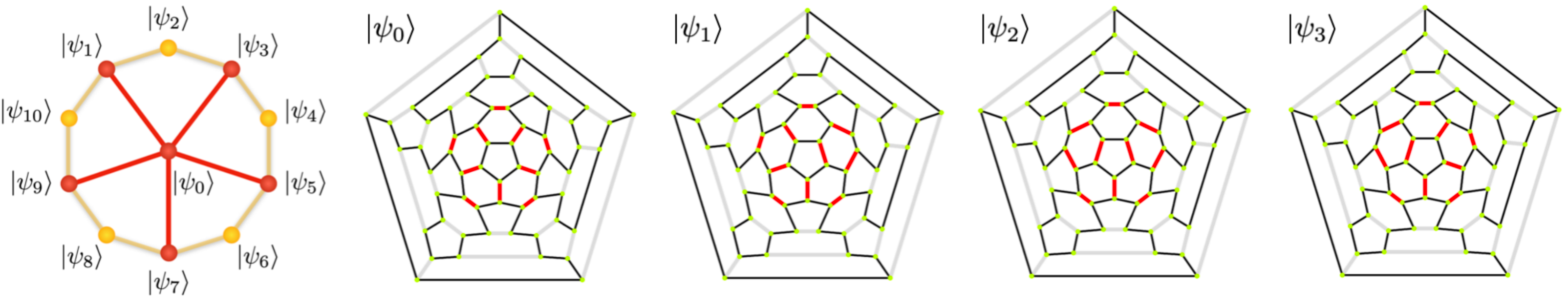}
\caption{Graphical representation of a self-contained sector within the Hilbert space. The nodes represent dimer covers, with four shown explicitly. A pair of nodes connected by a bond represents two dimer covers that are connected by a single hexagon-flip process.  For example, by flipping one hexagon at a time, we may traverse the following loop: $\vert \psi_0\rangle \rightarrow \vert \psi_1 \rangle \rightarrow \vert \psi_2 \rangle \rightarrow \vert \psi_3 \rangle \rightarrow \vert \psi_0\rangle$.  }
\label{fig.scar}
\end{figure*}

As seen from Fig.~\ref{fig.dos}, the density of states shows prominent peaks at several energies. These arise from small sectors of the Hilbert space, consisting of a few states that differ by local rearrangements. In each of these sectors, the dynamics is restricted to a small region within the C$_{60}$ graph, with the region outside remaining untouched. In this sense, the dynamics is non-ergodic. The resulting states resemble quantum scars\cite{Turner2018}, albeit in a finite system. We illustrate this phenomenon in Fig.~\ref{fig.scar} which depicts a sector consisting of eleven states (dimer covers). These states form a wheel-like structure with a central node and five spokes. The central node, $\vert \psi_0\rangle$, is a dimer cover with precisely five flippable hexagons, all immediately surrounding a pentagon. The states on the rim, labelled $\vert \psi_\ell \rangle$ with $\ell=1,\ldots, 10$, can be accessed from the central node by one flip or two flips. In all the dimer covers involved, the dynamics is restricted to the central region -- where the figures show dimers in red. These dimers are mobile as they are located on flippable hexagons. In contrast, grey bonds on the periphery remain static. 

Dynamics within this sector can be easily solved as a 11-site tight-binding problem. Analytic expressions can be found for the eigenvalues and eigenfunctions. When $V=0$, the eigenvalues (in units of $t$) are $\pm 2 \cos(2\pi/5) $, $\pm 2\cos(\pi/5)$, $\pm 3$ and $0$. The eigenvalues expressed using cosines are doubly degenerate, while the others are non-degenerate. In the density of states of Fig.~\ref{fig.dos}, we see sharp peaks at precisely these eigenvalues. The peaks arise due to high multiplicity, which can be seen as follows. The dynamic region in this sector in centred around a pentagon. We may take any one of the 12 pentagons of C$_{60}$ to be this central pentagon. In addition, for a given central pentagon, ten distinct arrangements are possible for the inert outer (grey) region. This leads to 120 copies of the 11-state sector of Fig.~\ref{fig.scar} in the Hilbert space. This phenomenon underlies the occurrence of spikes in the density of states of Fig.~\ref{fig.dos}.

These scar-like states are a fairly generic phenomenon. Their high multiplicity survives for $V\neq 0$, as the structure of each sector remains the same. While the energy eigenvalues may change with $V$, the multiplicity of eigenvalues remains intact. Sectors with the same structure can also appear on larger graphs. For example, this structure can exist in larger fullerenes such as C$_{70}$. It can also appear on larger graphs that describe Goldberg polyhedra\cite{Goldberg1937,Hart2013}.

\section{Ground state properties}
\label{sec.gs}
Below, we discuss ground state character in the `undoped' model with the full C$_{60}$ graph shown in Fig.~\ref{fig.c20_schlegel}. We then discuss the effect of introducing a pair of vacancies.

\subsection{Ground states in the undoped model}
We diagonalize the 12500$\times$ 12500 Hamiltonian matrix of the C$_{60}$ QDM and identify the ground state. We restrict our attention to $0<V/t <1$. In this regime, the ground state falls within the maximally flippable sector which consists of 5828 dimer covers.

\begin{figure}
\includegraphics[width=3.1in]{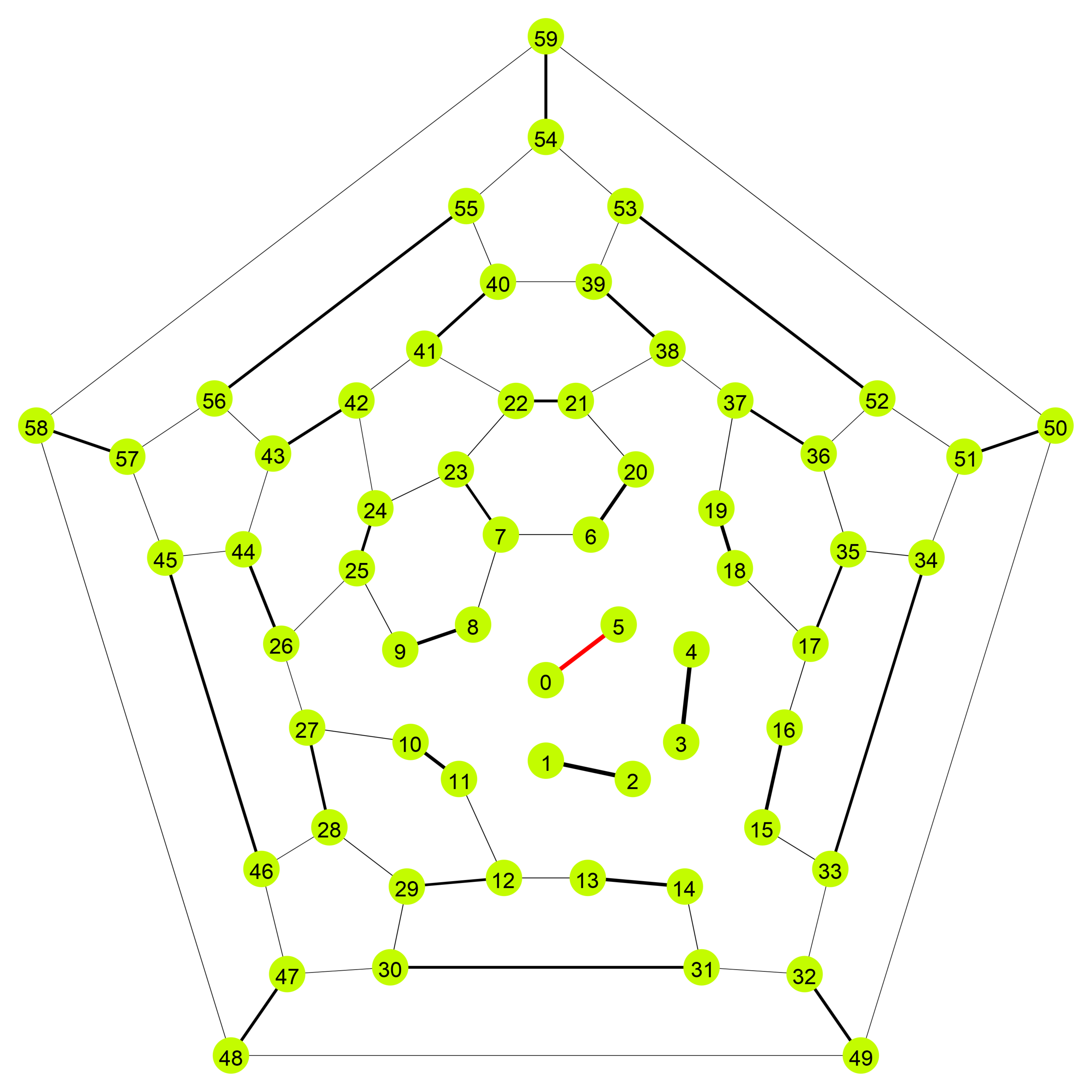}
\includegraphics[width=3.1in]{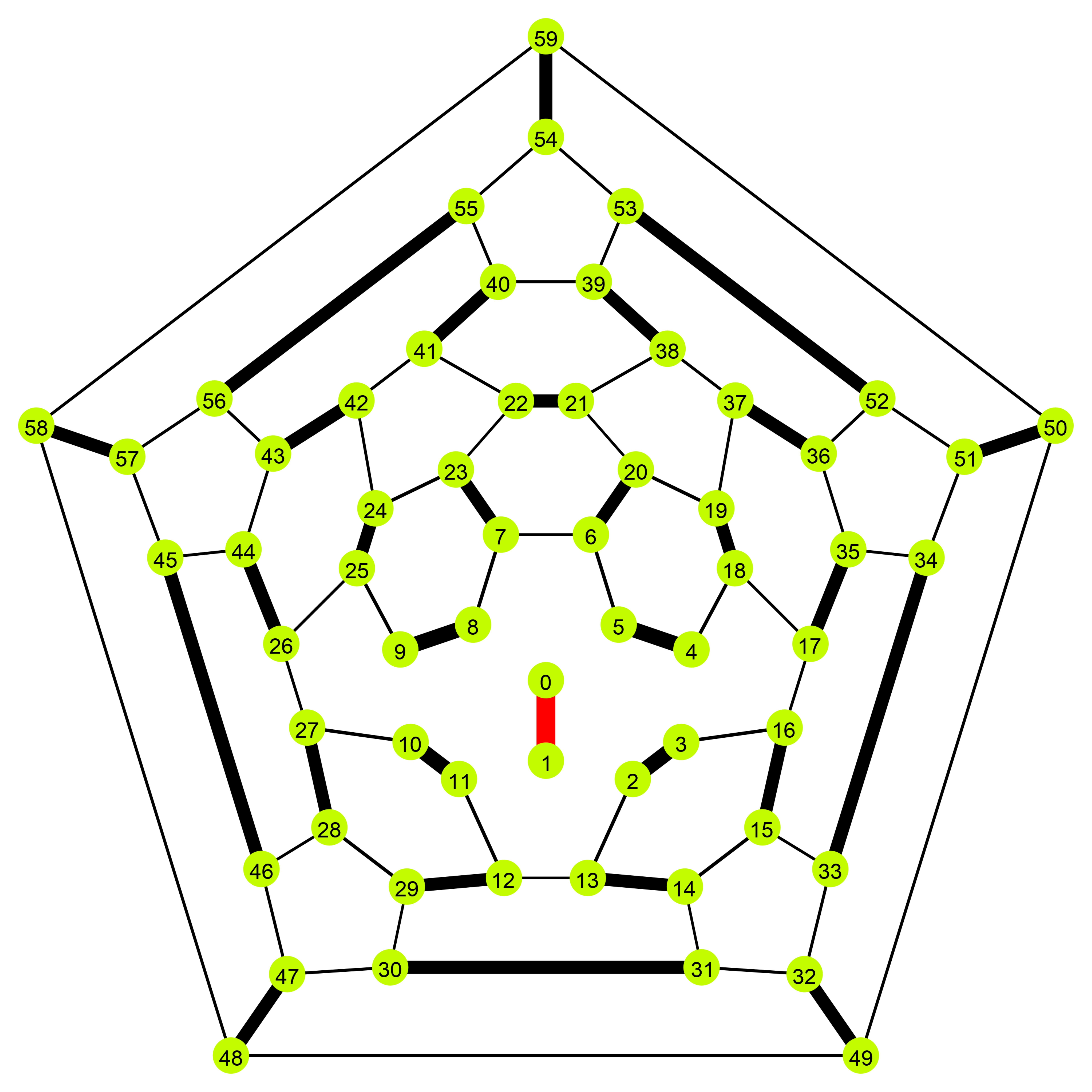}
\caption{Dimer correlations in the ground state of the C$_{60}$ QDM with $V/t=0$. The top and bottom panel correspond to two choices for the reference bond. In the top figure, the reference bond separates a pentagon from a hexagon. In the bottom figure, it separates two hexagons.}
\label{fig.C60corr}
\end{figure}

To describe the ground state, we may first consider the dimer-density on a given bond. Unlike C$_{20}$, the bonds in C$_{60}$ fall in two classes -- bonds that separate two hexagons and bonds that separate a hexagon from a pentagon. Within each class, the bonds are all equivalent. As we vary $V/t$, the dimer-density in the ground state is always higher on hexagon-hexagon bonds. Going further, we characterize the ground state in terms of correlations between bonds. This is shown in Fig.~\ref{fig.C60corr} for the case $V/t = 0$. The bond widths represent a joint probability -- the likelihood that a given bond as well as the reference bond are occupied by dimers. The reference bond may lie in either class. These cases are shown as two plots in the figure. We see strong correlations that span the entire C$_{60}$ graph. The pattern of correlations reflects the maximal flippable configuration shown in Fig.~\ref{fig.C60flippable}(bottom). Indeed, this configuration is the dominant component in the ground state. Its amplitude ranges from $\sim 0.1253 - 0.0134$, varying as $V/t$ is tuned from $0$ to $0.99$. In numerical terms, this may appear to be small. However, with 5828 dimer covers participating in the ground state, this represents a large contribution. The next largest contribution comes from 20 states that are connected to the maximally flippable configuration by a single flip. Each of these states carries an amplitude of $\sim 0.06695$ at $V/t=0$.

We may rationalize this finding as follows. In our regime of interest, $0< V/t <1$, the resonance kinetic energy dominates over the potential energy. The kinetic energy favours maximum spread of the wavefunction over the Hilbert space. To achieve this, the ground state chooses the largest sector which contains 5828 states. This sector can be viewed as a network where the central node is the maximally flippable configuration of Fig.~\ref{fig.C60flippable}(bottom). We reach other states by progressively flipping hexagons that are flippable. With each flip, we move away from the central node following various branches. To have maximal spread within this network structure, the wavefunction peaks at the central node. It explores every branch (and their sub-branches), with the amplitude decreasing steadily as we move away from the central node. This results in the maximally flippable configuration having the largest weight. 

If $V/t$ is increased beyond unity, there is a qualitative change, with the 256 unflippable states of the Hilbert space (e.g., Fig.~\ref{fig.C60flippable}(top)) having the lowest energy. As these states have no hexagons carrying three dimers, they minimize potential energy cost. 

\subsection{Ground states with doping}
We next consider `doping' the C$_{60}$ system with two vacancies. As the two vacancies can be positioned in various ways, we consider every possible relative position. The allowed states contain 29 dimers with no dimer touching a vacancy. We enumerate all such dimer covers using the FKT algorithm and a stochastic branching algorithm. For example, if the vacancies lie at the end of a bond that separates two hexagons, the resulting graph allows for $5500$ dimer covers. The resulting $5500\times5500$ QDM Hamiltonian matrix can be diagonalized to find the ground state energy. 

\begin{figure*}
\includegraphics[width=5.1in]{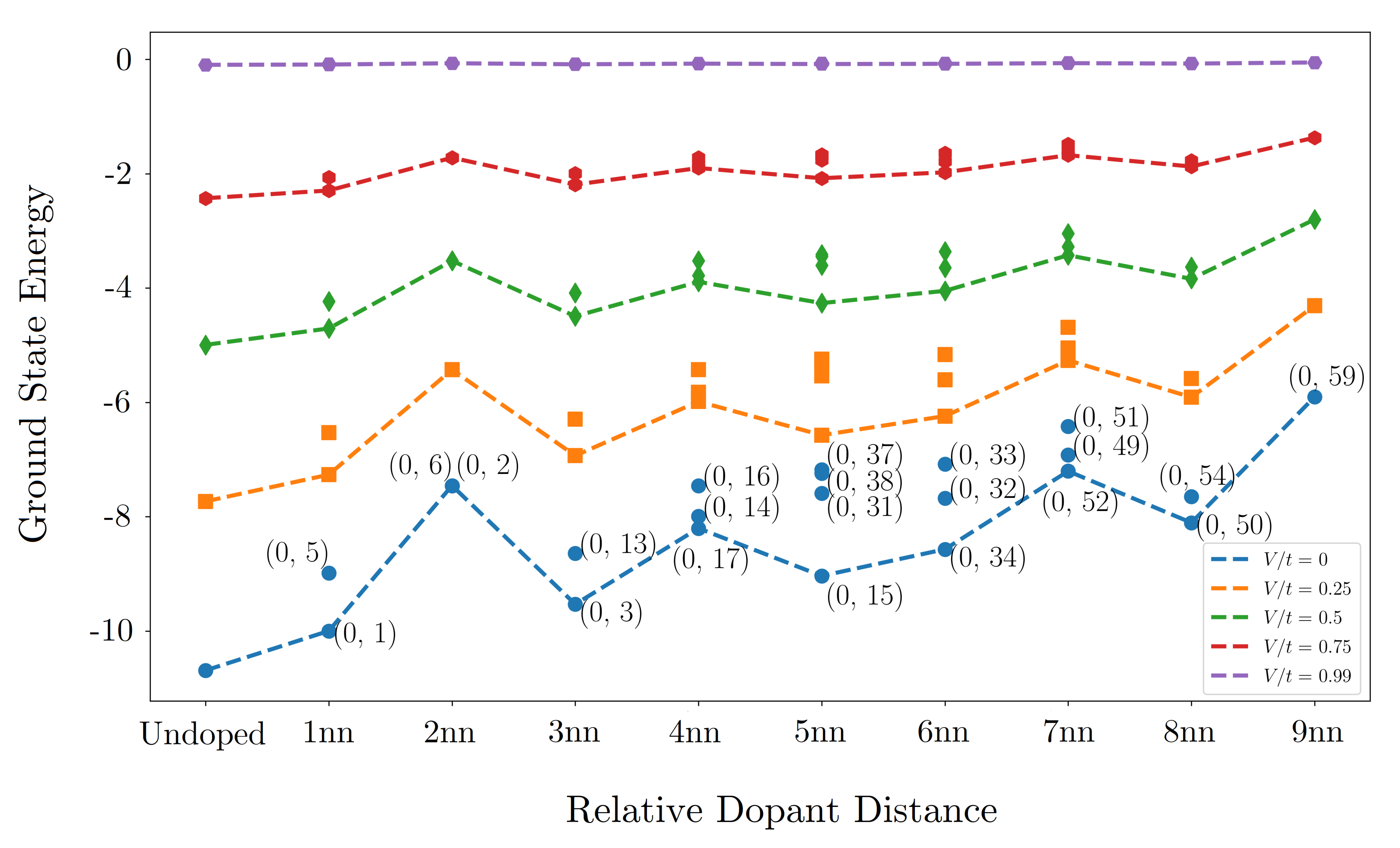}
\caption{Ground state energy in C$_{60}$ with two vacancies. The first column (`Undoped') gives the ground state energy with no vacancies, for various values of $V/t$. Subsequent columns show the ground state energy for various relative positions of two vacancies. For example, `1nn' corresponds to two positions that are nearest neighbours on the C$_{60}$ graph. There are two inequivalent choices for nearest neighbours, shown as (0,1) and (0,5). The positions are shown using the site labels given in Fig.~\ref{fig.c20_schlegel}. }
\label{fig.C60doped}
\end{figure*}

Fig.~\ref{fig.C60doped} shows the ground state energy for various relative positions of vacancies, as well as for various values of $V/t$. For all $V/t < 1$, the `undoped' case (with no vacancies) has lower energy than the doped case. In order to introduce vacancies, we must pay an energy cost (apart from any inherent binding energy). This cost stems from the kinetic energy of resonance. This can be seen, for example, as a reduction in the number of dimer covers that contribute to the ground state. In the `undoped' case, we have 5828 dimer covers that participate in the ground state. However, with two vacancies, we have $3064$ or fewer participating states.

Fig.~\ref{fig.C60doped} also shows the variation in ground state energy with separation between vacancies. For small $V/t$, the ground state energy is lowest when the vacancies are immediately adjacent to one another. As the vacancies move apart, the energy changes non-monotonically. However, there is a general upward trend as energy increases with separation. We interpret this as resonance-induced binding of vacancies. Each vacancy prohibits the presence of dimers in its immediate vicinity. This reduces the accessible space that dimers can explore -- effectively limiting their kinetic energy. With two vacancies, we have two such exclusion zones. If the two dimers are close to one another, the exclusion zones overlap, leaving a larger region accessible to resonance. This allows for maximal lowering of kinetic energy. 

We note that there are two inequivalent ways in which two vacancies can be adjacent -- they can lie at the ends of a pentagon-hexagon-bond or a hexagon-hexagon bond. With the former choice, three hexagons become unflippable. In contrast, the latter only prohibits dynamics on two hexagons. This is consistent with our finding that the latter has lower ground state energy (for small $V/t$).

At large $V/t$, where the potential energy dominates, energy does not vary with separation. The system tries to avoid three-dimer-hexagons as they contribute to a positive repulsion cost. For any positioning of vacancies, the optimal placement of dimers corresponds to a potential energy of zero.

\section{Discussion}

We model resonance processes in two fullerenes, C$_{20}$ and C$_{60}$, using a quantum dimer model description. In both cases, we find large quantum superpositions as ground states. These states can be viewed as mesoscopic spin liquids -- an intermediate case between benzene and lattice-spin-liquids.

We have used a quantum dimer model, implicitly assuming that distinct dimer covers are orthogonal. We believe this is a good approximation that does not affect results qualitatively. This can be seen by comparing our results with Ref.~\onlinecite{Flock1998}, which solves the Heisenberg model on C$_{60}$ by exact diagonalization in the nearest-neighbour-valence-bond basis. This rigorous approach leads to a ground state that is dominated ($\sim$ 99.82\%) by states of the maximally flippable sector discussed in Sec.~\ref{sec.maxflip} above. For comparison, the QDM yields a ground state that lies entirely within the maximally flippable sector. Within Heisenberg-like spin models, there is a well-known prescription to calculate overlaps of dimer covers. On the C$_{60}$ graph, no pair of dimer covers is orthogonal\cite{Flock1998}. However, the set of dimer covers is linearly independent\cite{Flock1998,Wildeboer2011}. This property provides some justification for using the QDM approach.

Previous theoretical studies on the fullerenes have used microscopic approaches such as the Hubbard model, the t-J model and the Heisenberg model. For smaller fullerenes, exact diagonalization has been used to solve these models\cite{Lin2007,Konstantinidis2005}. In the larger C$_{60}$ system, classical variational wavefunctions\cite{Coffey1992}, variational Monte Carlo\cite{Sheng1994,Krivnov1994} and density-matrix renormalization group (DMRG)\cite{Rausch2021} have been employed. All studies, including our QDM calculation (see Fig.~\ref{fig.C60corr}), obtain similar correlations in the C$_{60}$ ground state -- with the strongest correlation on hexagon-hexagon bonds.

QDMs and other models of constrained dynamics have been shown to host zero modes and scar states\cite{sen2022}. The fullerene QDM spectrum possesses a large number of zero modes, some of which arise from unflippable dimer covers. In the purely kinetic limit, zero energy states are protected with a relatively-large lower bound on their number. An interesting future direction is to examine whether these features survive when resonance on larger loops is included.  Apart from zero modes, the spectrum has many scar-like states which manifest as peaks in the density of states. They represent localized dynamics. Their high multiplicity can provide signatures in spectroscopic measurements. Similar features may also be found on larger systems, e.g., on Goldberg polyhedra\cite{Goldberg1937,Hart2013}.

We have demonstrated resonance-induced binding of vacancies within the QDM. Physically, vacancies can be introduced on the fullerene cage in at least two ways. One is through substitution, exemplified by dihydrogenated fullerene, C$_{60}$H$_2$. The Hydrogen atoms can be added at various relative positions. For any choice, they saturate two of the Carbons on the cage, disallowing their participation in dimer formation. Quantum chemical calculations of Ref.~\onlinecite{Matsuzawa1992} show that the most stable configuration has two Hydrogen atoms on neighbouring sites that are connected by a hexagon-hexagon bond. Our QDM approach arrives at the same conclusion, using a simple picture for the bonding processes involved. A second approach is to induce ionic character in the C$_{60}$ cage\cite{Krishnamurthy1992,Wabra2019}. It has been suggested that molecular solids of the form A$_3$C$_{60}$, where A is an alkali metal, may have C$_{60}$ in the $2-$ ionic state\cite{Baskaran1991,Baskaran2017}. In this point of view, C$_{60}^{2-}$ has higher stability due to resonance, driving superconductivity in these materials. Our QDM provides some support for this picture, with a direct explanation for resonance-induced correlations within a single fullerene molecule.

Resonating valence bond (RVB) theory has been invoked to explain the occurrence of superconductivity in the fullerenes\cite{Baskaran1991,Chakravarty1991,Chakravarty1991Science,Baskaran2017}. These studies are formulated at the mean-field level, presupposing a spinon-holon description. The QDM description shows RVB ideas in more direct fashion -- by describing resonance in a manner that can easily visualized. Exciting future directions include incorporating dynamics of vacancies and dimer-hopping between adjacent fullerene cages.

\acknowledgments
We thank G. Baskaran for insightful discussions. RKP thanks the International Centre for Theoretical Sciences (ICTS-TIFR) for hospitality and support.

\bibliographystyle{apsrev4-1} 
\bibliography{fullerene}
\end{document}